\documentclass[nofootinbib,twocolumn,pre,superscriptaddress]{revtex4}
\usepackage{tikz}
\usepackage{graphicx}%
\usepackage{color} 
\usepackage{transparent}
\usepackage{psfrag}
\usepackage{dcolumn}
\usepackage{amsmath}
\usepackage{amssymb}
\usepackage{latexsym}
\usepackage{hyperref} 
\usepackage{booktabs}
\usepackage{latexsym}
\usepackage{undertilde}

\begin{document}

\newcommand{\tikzcircle}[2][red,fill=red]{\tikz[baseline=-0.5ex]\draw[#1,radius=#2] (0,0) circle ;}%
\def\bea{\begin{eqnarray}}
\def\eea{\end{eqnarray}}
\def\beq{\begin{equation}}
\def\eeq{\end{equation}}
\def\f{\frac}
\def\k{\kappa}
\def\e{\epsilon}
\def\ve{\varepsilon}
\def\be{\beta}
\def\D{\Delta}
\def\h{\theta}
\def\t{\tau}
\def\a{\alpha}

\def\cDa{{\cal D}[X]}
\def\cD{{\cal D}[x]}
\def\cL{{\cal L}}
\def\cLo{{\cal L}_0}
\def\cLa{{\cal L}_1}

\def\Re{{\rm Re}}
\def\sj{\sum_{j=1}^2}
\def\rk{\rho^{ (k) }}
\def\rek{\rho^{ (1) }}
\def\cek{C^{ (1) }}
\def\rz{\rho^{ (0) }}
\def\rt{\rho^{ (2) }}
\def\rtb{\bar \rho^{ (2) }}
\def\trk{\tilde\rho^{ (k) }}
\def\trek{\tilde\rho^{ (1) }}
\def\trz{\tilde\rho^{ (0) }}
\def\trt{\tilde\rho^{ (2) }}
\def\r{\rho}
\def\tD{\tilde {D}}

\def\s{\sigma}
\def\kb{k_B}
\def\bF{\bar{\cal F}}
\def\F{{\cal F}}
\def\la{\langle}
\def\ra{\rangle}
\def\nn{\nonumber}
\def\up{\uparrow}
\def\dn{\downarrow}
\def\S{\Sigma}
\def\dg{\dagger}
\def\d{\delta}
\def\p{\partial}
\def\l{\lambda}
\def\L{\Lambda}
\def\G{\Gamma}
\def\o{\Omega}
\def\w{\omega}
\def\g{\gamma}

\def\jv{ {\bf j}}
\def\jr{ {\bf j}_r}
\def\jd{ {\bf j}_d}
\def\jdd{ { j}_d}
\def\noi{\noindent}
\def\a{\alpha}
\def\d{\delta}
\def\p{\partial} 

\def\la{\langle}
\def\ra{\rangle}
\def\e{\epsilon}
\def\n{\eta}
\def\g{\gamma}
\def\break#1{\pagebreak \vspace*{#1}}
\def\hf{\frac{1}{2}}

\title{Bidirectional motion of filaments: Role of motor proteins and passive cross linkers}
\author{Subhadip Ghosh}
\affiliation{Institute of Physics, Sachivalaya Marg, Bhubaneswar 751005, India
}
\affiliation{Homi Bhaba National Institute, Anushaktigar, Mumbai 400094, India}
\author{V N S Pradeep}
\affiliation{Indian Institute of Technology Hyderabad,
Sangareddy  502285, 
India
}
\author{Sudipto Muhuri}
\affiliation{Department of Physics, Savitribai Phule Pune University, Ganeshkhind, Pune 411007, India
}
\author{Ignacio Pagonabarraga}
\affiliation{Departament de F\'{\i}sica de la Mat\`eria Condensada, Universitat de Barcelona, Carrer Mart\'{i} i Franques 1, ES 08028 Barcelona, Spain}
\author{Debasish Chaudhuri}
\email{debc@iopb.res.in}
\affiliation{Institute of Physics, Sachivalaya Marg, Bhubaneswar 751005, India
}
\affiliation{Homi Bhaba National Institute, Anushaktigar, Mumbai 400094, India}


\begin{abstract}
 In eukaryotic cells, motor proteins (MP) bind to cytoskeletal filaments and move along them in a directed manner generating active stresses. During cell division a spindle structure of overlapping antiparallel microtubules (MT) form whose stability and dynamics under the influence of MPs has been  studied extensively. Although passive cross linkers (PCL) were known to provide structural stability to filamentous network, consequences of the interplay between ATP dependent {\em active} forces of MPs and {\em passive }entropic forces of PCLs on MT overlap remains largely unexplored. Here, we formulate and characterize a model to study this, using linear stability analysis and numerical integration.  In presence of PCLs, we find dynamic phase transitions with changing activity exhibiting regimes of stable partial overlap with or without oscillations, instability towards complete overlap, and stable limit cycle oscillations that {emerge} via a supercritical Hopf bifurcation characterized by an oscillation frequency determined by the MP and PCL parameters.  We show that the overlap dynamics and stability {depend} crucially on whether both the MTs of overlapping pair are movable or one is immobilized, having potential implications for {\em in vivo} and {\em in vitro} studies. 
\end{abstract}

\maketitle

\section{Introduction}

The cytoskeleton in eukaryotic cells is a complex network of interlinking filaments associated with motor proteins (MP) and passive cross linkers (PCL). It plays a central role in controlling  the cell shape and motility apart from being also involved in intra-cellular transport.
For example, cytoskeleton controls mechanical processes underlying the essential mechanism of cell division. In fact, the process of chromosome segregation  to daughter cells in eukaryotes is achieved by the formation of spindle structure comprising of overlapping microtubules (MTs)~\cite{Alberts2007,Wittmann2001,Walczak2008}. In a spindle, {\em interpolar} MTs from the two microtubule originating centers (MTOC) grow antiparallel to each other to interdigitate at the cell midzone. 
On the other hand, {\em astral} MTs  grow towards the cell membrane, where active force generators like Dynein capture these MTs and pull them towards the cell cortex~\cite{Grill2003}. 
The Kinesin-8 family of motor proteins (MPs) present at the interdigitating midzone control MT length~\cite{Fukuda2014}, and generate sliding force reducing MT overlap~\cite{Grill2003,Grill2005,Pecreaux2006,Su2013}. 
The deatails of the resultant dynamics has been thoroughly studied during the last decade~\cite{Gross2002,Kural2005, Levi2006,Leduc2010,Pavin2012,Chaudhuri2016,McLaughlin2016}.  
 
A large body of experimental and theoretical work focussed on active MP-filament systems~\cite{Holzbaur2010}, e.g., actin- myosin complex that forms the cell cortex~\cite{Tolic-Norrelykke2004,Brugues2012}, and MP driven active self assembly of filamentous proteins~\cite{Lam2014,Lam2016}. {Theoretical and experimental studies of spindle dynamics, and spindle oscillation, mainly focused on two aspects -- active force generation due to MPs and dynamic instability of the microtubules~\cite{Grill2005,Muhuri2011,Malgaretti2016,Ma2014,Pavin2012,Leduc2010,Scharrel2014}. }
These studies have clarified the subtle and rich interplay between MTs and MPs in the emergence and stability of the cytoskeleton, as well as qualitative differences of active complexes with respect to  their equilibrium counterparts.
Although, presence of  PCLs in the cytoskeleton has been known for a long time, relatively less attention has been paid to their role in determining the various structural properties of cytoskeletal network. Only recently it has become clear that they may also play important role in the spindle formation during mitosis. 
For example, it has been shown that PCLs can increase sliding viscous friction between overlapping MTs~\cite{Tawada1991}, and provide stability to the overlap~\cite{Braun2011}. In fact, it has been predicted that PCLs can stabilize the overlap of antiparallel MTs in the mitotic spindle, competing against MP pull~\cite{Johann2015, Johann2016}.   Recent studies on {\em in vitro} motility assays have identified  the entropic force  induced by PCLs as a relevant physical mechanism contributing to the stability of overlapping MT parallel arrays~\cite{Walcott2010, Lansky2015}. Nonetheless, the dynamic interplay between PCLs that promote MT overlap and MPs that pull them away from each other, and their impact in the self-assembly and stability of MP and PCL complexes {remain} largely unexplored.

In this paper, we present a comprehensive study of the non-equilibrium phases that develop as a result of the competition between the active forces generated by MPs and the passive forces due to PCLs. We consider a simplified geometry, where two anti-parallel,  length-stabilized MTs are pushed inwards to larger overlap by PCLs, while MPs pull them apart. This geometry, reminiscent of the MT arrangement in  the mitotic spindle, allows us to identify the fundamental principles that determine the impact that the dynamic interaction between MPs and PCLs has on the MT morphology. Specifically, we will focus on two complementary  arrangements and will  compare the behavior of two movable MTs with the situation in which one of the two MTs is  surface immobilized. Our study shows the existence of a variety of non-equilibrium phases, comprising of stable and unstable  overlapping arrays, as well as the  emergence of persistent characteristic oscillations. The development of such oscillations can be understood in terms of a stable limit cycle arising via a supercritical Hopf bifurcation.  The general framework identifies the relevant parameters that compete  and control the relative stability of these different phases. For example, we  show that the arrangements can be stabilized by tuning the MP {activity}, which itself depends on the  ATP concentration in the medium. Further, the stabilization is strongly dependent on whether one partner MT is immobilized, which could emerge in the cell via formation of strong bonds to the cytoskeletal matrix. {The theory also} provides an illustration of how the cell can control its dynamic arrangements by regulating the ATP concentration and the amount of passive cross linking. Our detailed predictions may be verified in proposed {\em in vitro} experiments, while many other qualitative aspects may be tested across {\em in vivo} studies.

\section{Model}
We consider two simple geometric arrangements comprising of a pair of antiparallel, {overlapping MTs} in presence of MPs and PCLs. In the first arrangement, we consider a situation in which one of the two overlapping MTs is made immobilized by fixing it to a substrate. The other MT can move freely under the entropic force of PCLs and active force due to MPs, one end of which are attached irreversibly to the substrate while the other end move unidirectionally on the MT as depicted in Fig.\ref{fig:1MT}.  In the second arrangement, we take up a situation { where both the overlapping MTs} are free to move and are subjected to active and passive forces due to MPs and PCLs, respectively, as depicted in Fig.\ref{fig:ful_sys_mod}. The active forces exerted by MPs are controlled and characterized by the ambient ATP concentration, and the concentration of MPs, $\phi_{MP}$. The stability of overlapping morphology is determined by the MP dynamics and its competition with the entropic forces exerted by $n_p$ number of PCLs diffusing within the overlap region.

Building on well-established approaches to the theoretical study of  filament- motor protein complexes~\cite{Grill2005}, 
MPs are modeled as active stretchable springs undergoing active attachment- detachment dynamics, governed by energy consumption via ATP hydrolysis, and {do} not obey detailed balance.  On an average, the maximum number of motor proteins that can attach to a MT is $N=L\, \phi_{MP}$, where $L$ denotes the length of the MT. Within mean field approximation, the kinetics of $n_m$ attached MPs can be expressed as   
\begin{equation}
\frac{dn_m}{dt} = \omega_a (N-n_m) - \omega_d n_m\exp\left[\frac{|f_l|}{f_d}\right],
\label{Eq:att_det_dyn_motor1}
\end{equation}
where $\w_a$ and  $\w_d$ denote the attachment and {\em bare} detachment rates respectively. The actual detachment rate of MPs increases exponentially with the load they are subjected to, $|f_l|$, irrespective of whether the load is extensile or compressive, and the detachment force $f_d$ sets its force scale. Modeling the MP as a linear spring of spring constant $k_m$, its extension $y^i$ leads to the load force felt by it $f_l = k_m y^i$.  

The ATP dependent activity allows attached MPs to walk along a filament, with a speed that depends on the force they are subjected to, $v_m(f_l)$, toward one of the filament ends. 
According to the experimental evidence  for  a variety of MPs, including kinesin~\cite{Carter2005}, we consider  a piecewise linear force-velocity relation
\[
v_m(f_l) =
\left\{
\begin{array}{ll}
v_0 & \mbox{for}\;\;\; f_l \leq 0 \\
v_0  \left( 1-\frac{f_l}{f_s} \right)& \mbox{for}\;\;\; 0<f_l \leq f_b,\, f_b>f_s  \\
-v_{back}& \mbox{for}\;\;\; f_l >f_b, \\
\end{array}
\right.
\]
where, $f_s$ denotes the stall force and $v_0$ stands for the intrinsic MP velocity.  For a load force beyond stall,  $f_l \geq f_b > f_s$,  the velocity saturates to an extremely small negative value $v_{back} = v_0\,(f_b-f_s)/f_s$~\cite{Leduc2010,Carter2005}, while  supportive loads do not affect the intrinsic MP motion.  All the parameters $v_0$, $\w_a$, $\w_d$, $f_s$, and  $f_d$ characterizing MPs are potentially functions of the ATP concentration in the ambient fluid. An assumption of Michaelis-Menten kinetics of ATP hydrolysis  successfully describes the  ATP dependence of $v_0$ for  kinesin, where $v_0$ increases with ATP concentrations to eventually saturate~\cite{Schnitzer2000}. This large variability of $v_0$ leads to various interesting dynamical regimes. We return to this point later in the paper. 

Neglecting fluctuations, the mean extension of $i$-th MP $y_i$ attached to a MT is  determined by its active velocity $v_m$  and the variation of the MT- overlap $x$, 
\begin{equation}
\frac{dy^i}{dt} = v_m(k_m y^i) + \frac{dx}{dt}.
\end{equation}

Finally, the dynamics of the overlap length, $x$,  between the two antiparallel MTs is determined by the mechanical balance of the forces they are subjected to,
\bea
\g_f \frac{dx}{dt} = -\sum_{i=1}^{n_m} k_m y^i + F_p(n_p),
\label{eq:forcebalance}
\eea
where, the left hand side corresponds to the viscous friction force associated with the relative motion between the two MTs,  quantified by the viscous friction $\g_f$.  
This force is balanced by the  total restoring force of the MPs as they  distort when attached to a MT due to their inherent rectification, and the force exerted by PCLs, $F_p$,  that depends on the number of PCLs, $n_p$. Here, $ \sum_{i=1}^{n_m} k_m y^i $  denotes the total force exerted by the attached  MPs. We stick to the mean field assumption that the total reaction force is uniformly shared by all the attached MPs,  hence  $f_l = -\frac{1}{n_m}  \sum_{i=1}^{n_m} k_m y^i $,  an assumption valid for a rigid rod like MT~\cite{Badoual2002,Vilfan1998}. 

In the presence of $n_p$ number of diffusing PCLs cross-linking two MTs in the overlap region, an entropic force $F_p = \kb T n_p/x$ develops. This  one dimensional ideal gas pressure pushes the system to increase its overlap region~\cite{Lansky2015}. Moreover, PCLs {increase} the viscous friction associated with the relative motion of the MTs to $\g_f = \g \exp(\zeta n_p)$, see Appendix-A.

 In the mitotic spindle, both the force due to cortical dynein transmitted from {\em astral} MTs via the spindle pole body~\cite{Grill2003, Grill2005}, and force due to 
bipolar kinesins potentially present at the interdigitating {\em interpolar} MTs~\cite{Sharp1999},  slide MTs actively to reduce their overlap.
On the other hand passive cross linkers like Ase1 generate entropic {force} that has the opposite effect, and tends to increase the overlap between the MTs. We analyze in depth the interplay between these two mechanisms and the impact they have in the stability of overlapping MTs. We select two simplified, although experimentally realizable {{\em in vitro}} assays consisting of kinesins, MTs and PCLs, where active forces due to kinesins compete with entropic force due to passive PCLs. We are able to identify the relevant morphological structures for the corresponding overlapping MTs and describe a thorough study that provides the relevant phase diagrams for these systems.



\section{Microtubule conformations stabilized by PCLs}
\subsection{One movable MT}
\label{sec:model1}
\begin{figure}[t]
\begin{center}
\includegraphics[width=8cm]{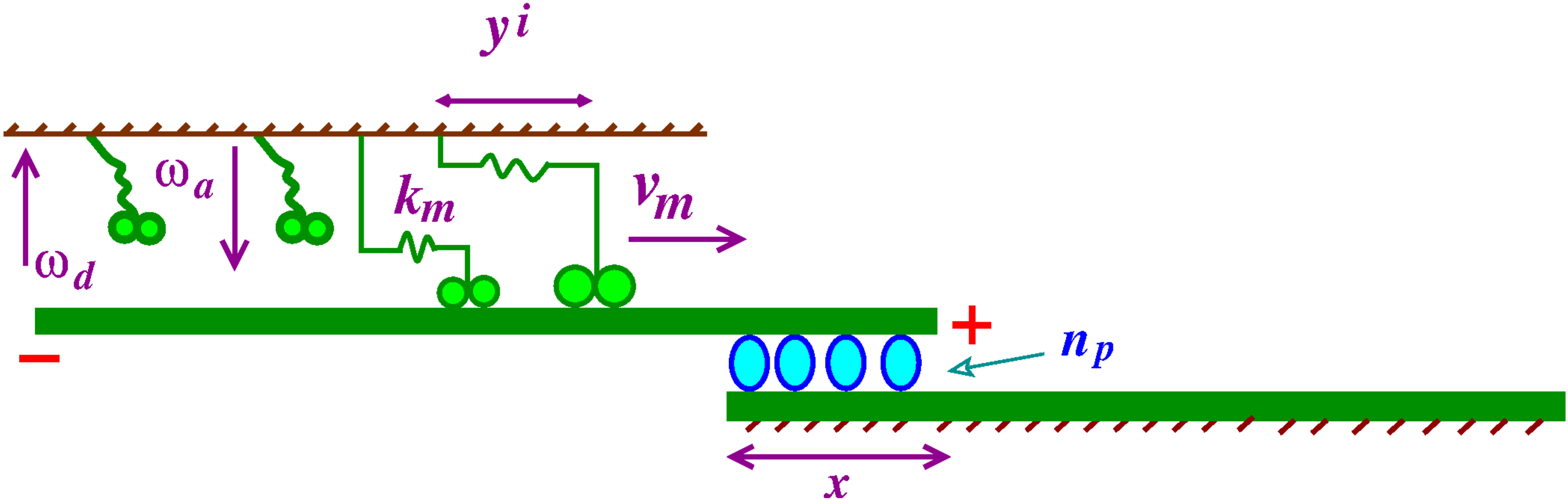}
\caption{(color online) 
Schematic diagram representing an {\em in vitro} experiment where a movable MT denoted by the green bar with $-$ and $+$ ends (left) slides over an immobilized MT attached to a surface (right). The attachment and detachment rates of MPs are denoted by $\w_a$ and $\w_d$ respectively.  When attached, $i$-th MP walks along the movable filament towards the plus end with a velocity $v_m$ that decreases with the extension $y_i$ generating a load. The overall effect of attached MPs is to pull the movable filament towards left reducing the overlap $x$. The overlap length contains $n_p$ number of PCLs that diffuses within it generating entropic force pushing the movable MT towards increasing $x$.}
\label{fig:1MT}
\end{center}
\end{figure}
%
Let us consider first a setup where one MT is attached irreversibly to a cover slip. This MT  is associated with a second, moving MT via PCLs, such as Ase1~\cite{Lansky2015}, as depicted in  Fig.\ref{fig:1MT}. One end of MPs, e.g., kinesins are irreversibly attached to the cover slip, while the other end stochastically bind and generate force on the movable MT. In the MT overlap region, PCLs  freely  diffuse, generating an entropic force $F_p = \kb T n_p/x$  tending to increase the overlap length, $x$, and increasing the effective viscous friction felt by the moving MT, $\g_f = \g \exp(\zeta n_p)$~\cite{Lansky2015}.  
The system evolves according to Eqs.(\ref{Eq:att_det_dyn_motor1})--(\ref{eq:forcebalance}), where we assume that after detachment the $i$ -th MP immediately relaxes to equilibrium with mean extension $y_i=0$. The average extension associated to  $n_m$ MPs is expressed as $y=\sum_i^{n_m} y^i/n_m$. 

Using the characteristic thermal energy $\kb T$, the time scale associated to MP detachment  rate $\omega_{d}^{-1}$,  and the typical length scale coming from force balance  $l_0 = \sqrt{\kb T/\g \w_d}$, the  dynamic equations, Eqs. (\ref{Eq:att_det_dyn_motor1})-(\ref{eq:forcebalance}) governing the motion of the overlapping MTs can be expressed in dimensionless form
\bea
\frac{d \tilde{x}}{d \tau} &=& \frac{1}{\exp(\zeta n_p)}\left[\frac{ n_p}{\tilde{x}}-N\tilde{n}_m \tilde{k}_m \tilde{y}\right], \nn\\
\frac{d \tilde{y}}{d \tau} &=& 
\tilde v_m(\tilde k_m \tilde y)+ \frac{d \tilde{x}}{d \tau}, \nn\\
\frac{d \tilde{n}_m}{d \tau} &=& (1-\tilde{n}_m)\tilde{\omega} -  \tilde{n}_m \exp\Big[\frac{\tilde{k}_m |\tilde{y}|}{\tilde{f}_d}\Big], 
\label{eq:dyn1}
\eea
where,  $\tau = t \omega_d$, $\tilde{x} = {x}/{l_0}$, $\tilde{y} = {y}/{l_0}$,  
$\tilde{n}_m = {n_m}/{N}$ the fraction of attached MPs,
$\tilde{k}_m = {k_m l_0}/{f}$, $\tilde{\omega} = {\omega_a}/{{\omega_d}}$, $\tilde{f}_s = {f_s}/{f}$, $\tilde{f}_d = {f_d}/{f}$, where 
$f = \sqrt{k_BT\gamma \w_d}$~\footnote{In dimensionless units, the MP velocity under a load reads
\unexpanded{\[
\tilde{v}_m =
\left\{
\begin{array}{ll}
 \tilde{v}_0 & \mbox{for}\;\;\; \tilde y\leq 0 \\
  \tilde{v}_0\left(1-\frac{\tilde{k}_m \tilde{y}}{\tilde{f}_s}\right) & \mbox{for} \;\;\;  0 < \tilde k_m \tilde y \leq \tilde f_b\\
  -\tilde{v}_{back} & \mbox{otherwise},
\end{array}
\right.
\]
}
where $\tilde{v}_0 = v_0/v$, $\tilde{v}_{back} = v_{back}/v$ with respect to the characteristic velocity  $v=l_0/\w_d$.
}. 
{
Table-\ref{table1} lists the used parameter values,  corresponding to kinesin at  room temperature. Under these conditions, the units of length, force and velocity are set by $l_0 = 33\,$nm,  $f= 0.125\,$pN, and $v= l_0 \w_d = 33\,$nm/s, respectively. 
\begin{table}[h]
\caption{Parameters: Two values of $v_0$ and $f_d$ correspond to ATP concentrations of $5\, \mu$M and 
2\,mM respectively.}
\centering
\begin{tabular}{l l l}
\hline\hline
active velocity & $v_0$ & 0.03, 0.8$\mu$m/s~\cite{Schnitzer2000,Block2003}\\
stall force & $f_s$ & 7.5 pN~\cite{Carter2005,Block2003} \\
back velocity & $v_{back}$ & 0.02$\mu$m/s~\cite{Carter2005} \\
detachment force & $f_d$ & 1.8, 2.4 pN\\
                             &          &  (Appendix-B)\\
attachment rate & $\w_a$ &  20/s~\cite{Schnitzer2000, Chaudhuri2016} \\
detachment rate & $\w_d$ & 1/s~\cite{Block2003} \\
motor stiffness & $k_m$ & 1.7 pN/nm~\cite{comment1}\\
ambient viscous friction & $\g$ & 893 $\kb$T-s/$\mu$m$^2$~\cite{Lansky2015} \\
friction parameter & $\zeta$ & 0.22 (dimensionless) \\
\hline\hline
\end{tabular}
\label{table1}
\end{table}%

\subsubsection{Linear stability of overlapping morphologies}
The  overlapping MTs steady configurations  are obtained from the fixed points of Eqs.(\ref{eq:dyn1}), which read as,
\begin{eqnarray}
 \tilde{y}_0 &=& \tilde{f}_s / \tilde{k}_m \nonumber\\
 \tilde{n}_m^0 &=& {\tilde{\omega}} / [{\tilde{\omega}+\exp({\tilde{f}_s}/{\tilde{f}_d})} ]\nonumber\\
  \tilde{x}_0 &=& { n_p}/ ({N \tilde n_m^0 \tilde{f}_s}),
\end{eqnarray}
and express the fact that the steady state emerges from  force balance, $n_m k_m y = n_p \kb T/x$. 
We can identify the stability of these overlapping morphologies analyzing the time evolution of  small perturbations around the fixed points of Eq.(\ref{eq:dyn1}), in terms of the MT overlap, MP extension and number of attached motors, $ | \psi \ra \equiv (\d \tilde x, \d \tilde y, \d \tilde n_m)^T$.~\footnote{Note that the expansion up to linear order around the fixed point, where load force $k_m y=  f_s$, requires that the force-velocity relation is continuous and differentiable at least once at this point. The piece-wise linear force-velocity relation used in this paper obeys this condition, with the discontinuities being at load force $=0$, and at $ f_b $ where $ f_b >  f_s$. 
The fact that kinesin has a small but finite back velocity has been observed experimentally~\cite{Carter2005} and allows $v_m(k_m y)$ to be analytic at $k_m y = f_s$. } 

To that end, we identify all the elements of the $3 \times 3$ stability  matrix {$\utilde {a}$} , that quantifies the evolution of these deviations,  $d | \psi \ra / d \t =$  {$\utilde {a}$} $ | \psi \ra$,  \begin{align}
& a_{11} = -{(N \tilde{n}_m^0 \tilde{f}_s)^2} / { n_p \exp(\zeta n_p)}, \,\,
a_{12} = -{\tilde{k}_m N \tilde{n}_m^0} / {\exp(\zeta n_p)}, \nn\\
& a_{13} = -{\tilde{f}_s N} / {\exp(\zeta n_p)}, \,\, a_{21} = a_{11}, \nn\\
& a_{22}= -\left(\frac{\tilde{v}_0 \tilde{k}_m}{\tilde{f}_s}+\frac{\tilde{k}_m N \tilde{n}_m^0}{\exp(\zeta n_p)}\right),\,\, a_{23} = a_{13},\,\, a_{31} = 0, \nn\\
& a_{32} = -\frac{\tilde{k}_m \tilde{n}_m^0}{\tilde{f}_d}\exp\left[ {\tilde{f}_s} /  {\tilde{f}_d}\right], \,\, a_{33} = -\left(\tilde{\omega}+\exp\left[{\tilde{f}_s} / {\tilde{f}_d}\right]\right) \nn. 
\end{align}
%
The steady configuration is stable when all the eigenvalues, $\lambda$ of $\utilde a$ have a real negative component. The eigenvalues are  obtained solving the characteristic equation 
\begin{eqnarray}
F(\l) = 
\lambda^3 + A \lambda^2 + B \lambda + C &=& 0,\label{Eq:cubic_act_pass01}
\end{eqnarray}
where, $A = - {\rm Tr}$(\,{$\utilde a$}\,), $B = \hf (a_{ii} a_{jj} - a_{ij} a_{ji})$ with $a_{ij}$ denoting elements of the matrix $\utilde a$, and 
$C = -$det(\,{$\utilde a$}\,).~\footnote{The full expressions for the coefficients of $F(\l)$ in terms of the elements of the stability matrix  $\utilde a$  read
\unexpanded{
\bea
A &=& \frac{(N \tilde{n}_m^0 \tilde{f}_s)^2}{ n_p \exp(\zeta n_p)} + \frac{\tilde{v}_0 \tilde{k}_m}{\tilde{f_s}}+\frac{N \tilde{n}_m^0 \tilde{k}_m}{\exp(\zeta n_p)} + \tilde{\omega}+\exp\left[\frac{\tilde{f}_s}{\tilde{f}_d}\right] \nn\\ \label{Eq:cubic_coeff_A}\\
B &=& \frac{(N \tilde{n}_m^0 \tilde{f}_s)^2}{ n_p \exp(\zeta n_p)}\left[\frac{\tilde{v}_0 \tilde{k}_m}{\tilde{f}_s}+\tilde{\omega}+\exp\left[\frac{\tilde{f}_s}{\tilde{f}_d}\right]\right] \nonumber\\&& + \frac{\tilde{v}_0 \tilde{k}_m}{\tilde{f}_s}\left[\tilde{\omega}+\exp\left[\frac{\tilde{f}_s}{\tilde{f}_d}\right]\right] \nonumber\\&& + \frac{N \tilde{n}_m^0 \tilde{k}_m}{\exp(\zeta n_p)}\left[\tilde{\omega}+\left(1-\frac{\tilde{f}_s}{\tilde{f}_d}
\right)\exp\left[\frac{\tilde{f}_s}{\tilde{f}_d}\right]\right] \label{Eq:cubic_coeff_B} \\ 
C &=&  \left(\frac{(N \tilde{n}_m^0 \tilde{f}_s)^2}{ n_p \exp(\zeta n_p)}\right)\left(\frac{\tilde{v}_0 \tilde{k}_m}{\tilde{f_s}}\right)\left(\tilde{\omega}+\exp\left[\frac{\tilde{f}_s}{\tilde{f}_d}\right]\right). \label{Eq:cubic_coeff_C}
\eea}
}
Since the active velocity $v_0 \geq 0$, one has  $A>0$, $C\geq 0$, 
while B may change its sign with changing $f_d$ or $f_s$. 
As a result, there are   only four possible combinations of roots of Eq.(\ref{Eq:cubic_act_pass01}), with one of them, $\l_1$, remaining always real and 
negative,  as is shown in the following subsection.  
Thus the long time dynamics is characterized essentially by the two other roots. This gives rise to four possible dynamical phases:
(i)~linearly stable ($s$): both these roots are real and negative $\l_{2,3} <0$ corresponding to a  stable node, (ii)~stable spiral ($ss$): $\l_{2,3} = \a \pm i \be$ with $\a<0$, 
(iii)~unstable spiral ($us$):  $\l_{2,3} = \a \pm i \be$ with $\a>0$, 
and (iv)~linearly unstable ($u$): both the other roots $\l_2, \l_3$ are real and positive giving rise to unstable nodes.  In general it is possible to have one positive and one negative real root corresponding to a saddle point, however, it is discarded by the form of coefficients in the polynomial $F(\l)$. 
A final possibility is when $\l_{2,3} = \pm i \be$ that occurs as $\a$ changes sign and reaches $\a=0$ denoting a {\em center}. As we will see later, this captures the onset of stable oscillations.  

All these phases characterize different types of overlap dynamics. Stability occurs when opposing forces due to active MPs and PCLs balance each other. In our analysis, instability is mediated by detachment of  all the MPs from the movable MT leading to its complete overlap with the immobile MT, driven entirely by the PCLs. Complex eignvalues denote development of oscillatory dynamics. We assume a constant number of PCLs in the overlap region, a reasonable choice given the long association time of PCLs to MTs at the overlap. At small overlap $x$, the entropic extensile force $\kb T n_p/ x$ dominates not allowing MT- overlap to vanish. Changing ATP concentration in the ambient medium changes activity parameters of MPs, as a result changing the effect of their competition with PCLs. We study the phase diagram in the plane of two such important active parameters, $\tilde v_0$ and $\tilde f_d$.

\subsubsection {The phase diagram}
The change in behavior of the eigenvalues allows us to build a phase diagram  that determines the expected MT configurations and dynamics. We use the fact that the characteristic polynomial in Eq.(\ref{Eq:cubic_act_pass01}) behaves asymptotically as $F(\l) \sim -|\l|^3$ for large negative $\l$ ensuring that it has to cross the $ F(\l)=0$ line at some negative $\l$ value. 
Thus one of the roots always remains real negative, $\l_1 <0$.
The phase diagram that identifies all possible configurations can be built once we identify the three relevant boundaires delimiting the change of  stability of the different regimes described in the previous subsection. Specifically, 

{\em a) Boundary between linearly (un)stable and  (un)stable spiral phases:}
The cubic polynomial in Eq.(\ref{Eq:cubic_act_pass01}) has a minimum at $\l_m = -\frac{A}{3} + \frac{1}{3}\sqrt{A^2 - 3B}$. 
For $B>0$  ($B<0$) the minimum occurs at negative (positive) $\l$.
At $F(\l_m)=0$, two degenerate negative (positive) real roots emerge, $\l_3=\l_2=\l_m$ making the dynamics linearly stable (unstable). 
On the other hand $F(\l_m)>0$  implies complex conjugate eigenvalues. The corresponding phase boundaries are given by (See Appendix-C) 
\begin{equation}
C = \bigg[\frac{1}{3}A + \frac{2}{3}\sqrt{A^2 -3B}\bigg]\bigg[-\frac{1}{3}A + \frac{1}{3}\sqrt{A^2 -3B}\bigg]^2,
\label{eq:pb_unStable}
\end{equation}
where $B>0$ ($B<0$) denotes the boundary between (un)stable spiral and  linearly (un)stable phase.

\begin{figure}[t]
\begin{center}
\includegraphics[width=8cm]{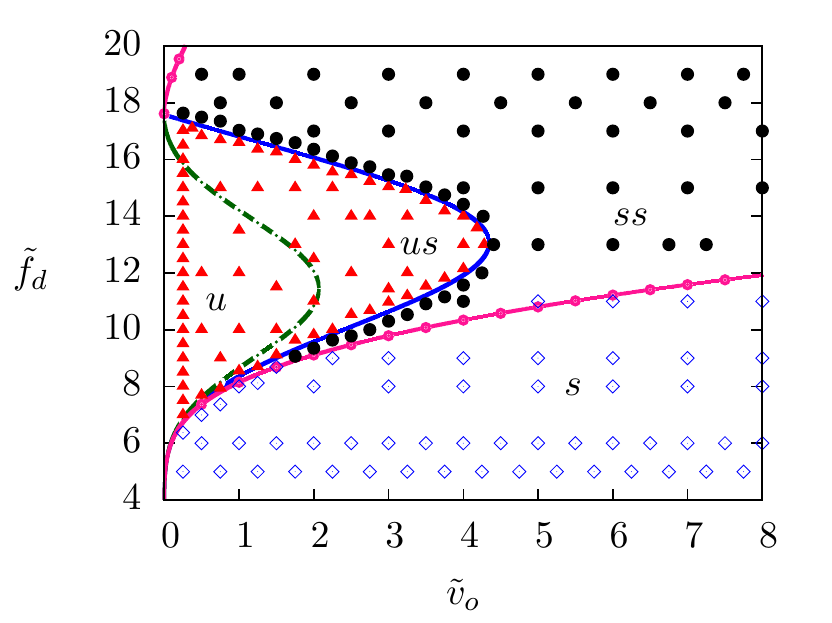}  
\caption{ (color online) 
Phase diagram of overlap with one fixed MT and one movable MT in $\tilde v_0-\tilde{f}_d$ plane. 
We used parameter values $\tilde k_m = 450$, $\tilde\omega = 20$, $\tilde{f}_s = 60$, $N = 5$ and $n_p = 15$. 
The points denote non-equilibriuam phases obtained from numerical solutions of full non-linear dynamical equations, Eq.(\ref{eq:dyn1}). We find the following phases:  (i)~stable {\color{blue} $\diamond$}, 
 (ii)~stable spiral \tikzcircle[black, fill=black]{2pt},  
 (iii)~stable limit cycle {\color{red} $\blacktriangle$}. The predictions of  linear stability analysis are indicated by symbols $s$: linearly stable, $u$: linearly unstable, $us$: unstable spiral, $ss$: stable spiral.
The lines denote phase boundaries calculated from Eq. (\ref{eq:pb_unStable}) -- the (green) dashed line is the boundary between $u$ and $us$, (blue) solid line is the boundary between $us$ and $ss$, and the (pink) beaded-line is the boundary between $s$ and $ss$. } 
\label{fig:act_pass_pb1}
\end{center}
\end{figure}

{\em b) Boundary between stable spiral and unstable spiral phases: } This phase boundary is characterized by the change in sign of the real part of complex conjugate roots which occurs at $\alpha = 0$. Hence, the cubic function in Eq.(\ref{Eq:cubic_act_pass01})  should read as $F(\lambda) = (\lambda + |\lambda_1|)(\lambda + i\beta)(\lambda - i\beta)$. Comparing coefficients of $\l$ with  Eq.(\ref{Eq:cubic_act_pass01}) we get $C = AB$ as the equation for this phase boundary, with a necessary condition that $B>0$. 
At this boundary, the stable spiral becomes unstable spiral, non-linear stabilization of whose oscillation- amplitude may give rise to a stable limit cycle. 

Fig.\ref{fig:act_pass_pb1} displays the phase diagram in the $\tilde v_0$-$\tilde f_d$ plane obtained both from the linear stability analysis and numerical solution of the model (a further detailed view of the phase diagram at a different $n_p,\,N$ value is depicted in Fig.~\ref{fig:phdia2}($a$) in Appendix-D). Continuous lines denote phase boundaries derived from linear stability analysis, while the points correspond to observed non-equilibrium morphologies  obtained  from numerical integration of the non-linear dynamics, Eq.~(\ref{eq:dyn1}). The non-linear dynamics shows existence of various steady states characterized by (i)~a stable phase with stable partial overlap of MTs, (ii)~a stable spiral phase, where the system shows decaying oscillation to a final stable state of partial overlap, and (iii)~a stable limit cycle phase with persistent oscillation of overlap.

\subsubsection{Stability, instability and limit cycle } 
We use Fig.~\ref{fig:time_evo} to analyze the overlap dynamics in detail. Deep inside the stable spiral phase denoted by $\tilde{f}_d = 16$, $\tilde{v}_0 = 3$, one obtains an inward spiral in the phase plot in fraction of bound MPs $\tilde n_m$ and overlap $\tilde x$ finally reaching a fixed point. This characterizes relaxation to a final stable state.  The time scale of the decay of oscillation amplitude gets longer with decreasing $\tilde{v}_0$, and finally the dynamics changes to small amplitude stable oscillation reached via a super-critical Hopf bifurcation~\cite{Strogatz2014}. 
This is characterized by a stable limit cycle surrounding an unstable fixed point in the $\tilde n_m - \tilde x$ plane, as shown in Fig.~\ref{fig:time_evo}($b$). Initial conditions inside or outside the limit cycle,  flow to the limit cycle via a spiral path in the phase space. The amplitude of this limit cycle oscillation  grows as one moves away from the phase boundary denoting the onset of a super-critical Hopf bifurcation. The values of $\tilde{f}_d$, $\tilde{v}_0$ used in Fig.~\ref{fig:time_evo}($a$) and ($b$) can be attained for wild type kinesins by changing ATP concentration, as is shown in Sec.~\ref{sec:expt}.


At large enough $\tilde v_0$, the dynamics of the average MP extension $\tilde y$ is essentially controlled by the force-velocity relation of MPs. A small increase of 
$\tilde y$ from the steady state is compensated by a negative response. As a result $\tilde n_m$ and $\tilde x$ remain stable. 

At intermediate values of  $\tilde v_0$,  a limit cycle oscillation develops due to non-linearity, cooperativity, and phase lag. 
The mechanism is elucidated in Fig.s~\ref{fig:time_evo}($c$)-$(e)$. 
Consider a state with the largest possible overlap $\tilde x$.  
The corresponding entropic force due to PCLs $\sim n_p/\tilde x$ is weak, thus MPs slide the movable MT to a smaller $\tilde x$. As a result, the mean extension $\tilde y$ relaxes to smaller values. This relies on the cooperativity of the MPs via load sharing. The force-velocity relation of MPs contain a finite $\tilde v_0$, that generate a phase lag between $\tilde x$ and $\tilde y$. This phase lag can easily be noticed comparing Fig.s~\ref{fig:time_evo}($c$) and $(d)$.
The decrease in $\tilde y$ reduces stress on the bound MPs leading to lesser detachment, and as a result an overall increase in the number of bound motors $\tilde n_m$. This mechanism keeps on reducing $\tilde y$ and increasing $\tilde n_m$ until the latter reaches its maximum. Due to the phase lag, even at this point overlap $\tilde x$ keeps on decreasing until reaching its own minimum.  
At this point $\tilde y$ starts to increase due to a finite $\tilde v_0$. This starts to decrease $\tilde n_m$ leading to an increase in $\tilde x$ that in turn increases $\tilde y$. This positive feedback leads to enhanced detachment of the bound motors. However, beyond a point the attachment rate $\w_a$ wins over the overall detachment rate as $\tilde n_m$ becomes too small. This is the point of minimum $\tilde n_m$ where $\tilde y$ reaches its largest value and overlap nears its peak. Then the cycle repeats. 
}

For vanishing $\tilde v_0$, the mean motor extension $\tilde y$ follows $\tilde x$, with negligible phase lag. Starting from a steady state, a slight increase in the extension $\tilde y$ decreases the number of attached MPs, $\tilde n_m$. This in turn increases the overlap $\tilde x$, which leads to further decrease in $\tilde n_m$. This mechanism underpins a run away instability, where the movable MT  increases indefinitely the extent of its overlap with the immobile MT.  

It is interesting to note that the time scales of growth or decay of perturbation near the phase boundaries show diverging behavior, reminiscent of equilibrium phase transitions (see the Appendix-E). 
\begin{figure}[t]
\begin{center}
\includegraphics[width=8cm]{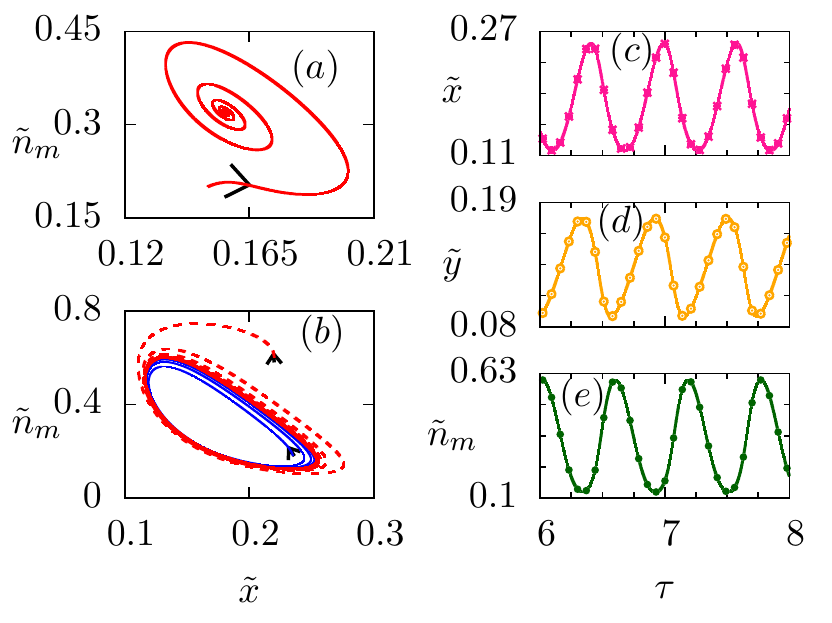} 
\caption{(color online) 
($a$)~Stable spiral: Phase space plot of the fraction of bound MPs $\tilde n_m$ and overlap $\tilde x$ at  $\tilde{f}_d = 16$ and $\tilde{v}_o = 3$, shows a stable spiral denoting decaying amplitude of oscillation to final stability.  The other parameters used  are the same as in Fig.~\ref{fig:act_pass_pb1}. 
($b$)~Stable limit cycle: The same phase space trajectory at $\tilde{f}_d = 16$ and $\tilde{v}_o = 2$ showing a stable limit cycle, with initial state inside (blue solid line)  or outside (red dashed line) spiraling towards it, denoting a super-critical Hopf bifurcation. This leads to stable oscillation of the overlap. 
The other panels show the corresponding evolution in  $\tilde x$ ($c$), $\tilde y$ ($d$) and $\tilde n_m$ ($e$) as a function of dimensionless time $\t$. }
\label{fig:time_evo}
\end{center}
\end{figure}

\begin{figure}[t]
\begin{center}
\includegraphics[width=6cm]{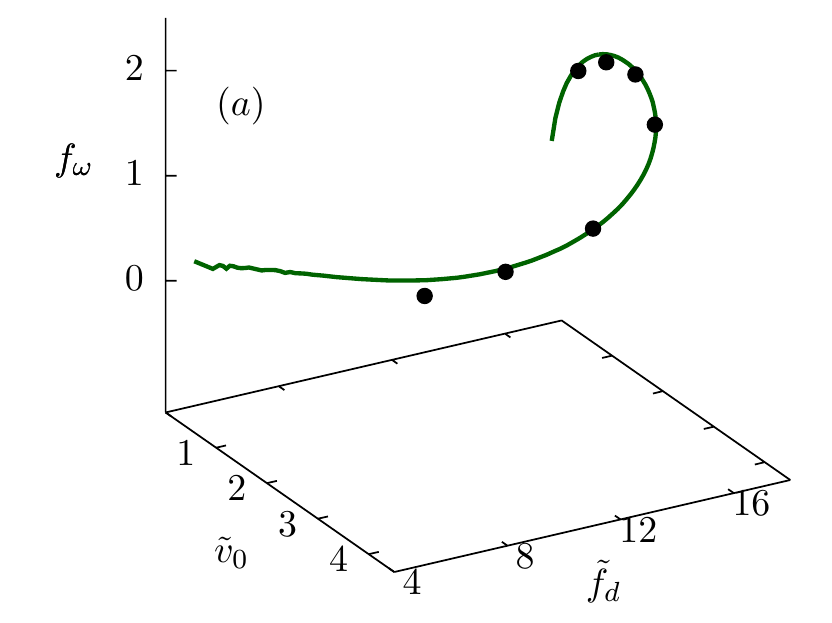} 
\includegraphics[width=8cm]{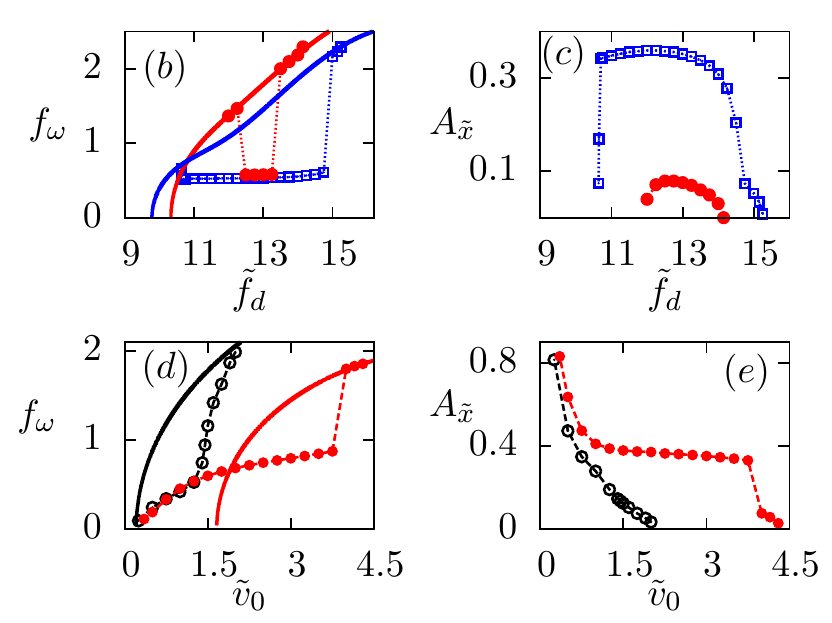}
\caption{(color online) 
($a$)~$f_{\omega}$~(Hz), the oscillation frequency of overlap length, along the supercritical Hopf-bifurcation line in Fig.(\ref{fig:act_pass_pb1}). The points denote numerical estimate, and the line is a plot of $f_\w=\sqrt{B}/2\pi$. The frequency changes by one order of magnitude from  0.2 Hz to 2 Hz.   
The other four panels show variation of  frequency and amplitude of overlap within the stable limit cycle phase  in Fig.(\ref{fig:act_pass_pb1}).
($b$) and ($c$) show this at fixed values of $\tilde v_0=$4 (red), 3 (blue) as $\tilde f_d$ is varied from the lower to upper Hopf bifurcation lines. 
$(d)$ and $(e)$ show the behavior with changing $\tilde v_0$, at fixed values of $\tilde f_d=$ 13 (red solid circle), 16 (black open circle). The solid lines in ($b$) and ($d$) denote $f_\w=\be/2\pi$ where $\be$ is obtained from numerical solution of Eq.(\ref{Eq:cubic_act_pass01}). This reduces to $f_\w = \sqrt{B}/2\pi$ at the supercritical Hopf bifurcation and captures the numerically obtained frequency from the non-linear dynamics near this phase boundary. Dashed lines are guide to eye. 
}
\label{fig:freq_plot1}
\end{center}
\end{figure}

\subsubsection{Stable limit cycles: Oscillation amplitude and frequency}
At the boundary between the stable spiral and unstable spiral phases, predicted by the linear stability analysis, two modes are  purely oscillatory, denoted by $\lambda_{2,3} = \pm i\beta$ (and $\l_1<0$).  Inserting this functional form in the secular determinant, Eq.(\ref{Eq:cubic_act_pass01})  gives $\beta = \sqrt{B}$.
 Thus, the frequency of oscillation characterizing the Hopf bifurcation boundary reads $f_{\omega} = \sqrt{B}/2\pi$, where $B$ is given by  Eq.(\ref{Eq:cubic_coeff_B}), which is sensitive to activity of MPs via $\tilde f_s$, $\tilde v_0$, $\tilde f_d$, $\tilde \w$, as well as the impact of PCLs through its number $n_p$ and modification of viscous friction $\g_f$.  
Fig.\ref{fig:freq_plot1}($a$) shows a good agreement between the expression $f_{\omega}$ with the numerically calculated limit cycle oscillation frequency along the super critical Hopf bifurcation boundary. Along this line, $f_{\w}$ varies between 0.2 to 2 Hz, a range easily observable in experimental assays. 

As we move away from the Hopf bifurcation boundary, the nature and properties of the frequency and amplitude of the stable limit cycle change substantially. For example, if we move in the phase diagram Fig.(\ref{fig:act_pass_pb1}) along a constant $\tilde v_0$ line from one Hopf bifurcation boundary to the other by varying $\tilde f_d$, the frequency of oscillation $f_{\omega}$ follows $\sqrt{B}/2\pi$ as one crosses the phase boundary from the stable spiral phase. However, inside the stable limit cycle $f_{\omega}$ drops to a relatively small value and remains almost independent of  $\tilde f_d$ (Fig.\ref{fig:freq_plot1}($b$)).  The amplitude of oscillation vanishes towards the phase boundary (Fig.\ref{fig:freq_plot1}($c$)), capturing a characteristic of supercritical Hopf bifurcation. 

On the other hand, if we move along a constant $\tilde f_d$ line reducing $\tilde v_0$ starting from a stable spiral phase, at the supercritical Hopf bifurcation again the frequency gets the form  $f_{\omega} = \sqrt{B}/2\pi$, however, it changes its functional dependence drastically as we enter the stable limit cycle phase where we observe a sharp decrease in the oscillation frequency associated with similar increase in amplitude (Fig.\ref{fig:freq_plot1}($d$), ($e$)).

\subsubsection{Re-entrant transitions:}
Increasing $\tilde f_d$ at a fixed value of $\tilde v_0$ it is possible to obtain re-entrant transitions from stable spiral - stable limit cycle - stable spiral phases, as shown in Fig.~\ref{fig:act_pass_pb1}. 
Staying within the stable limit cycle phase, one can go from the lower to the upper Hopf bifurcation boundary, varying $\tilde f_d$. The frequency and amplitude of oscillation show non-monotonic variation in general, with the frequency reaching a minimum and the amplitude a maximum in the middle of the limit cycle region. The phases below and above the limit cycle along the $\tilde f_d$ axis are both stable spiral. The small $\tilde f_d$  stable spiral region reaches a final stability with large overlap $\tilde x$, and as a result with little stored free energy in PCLs. On the other hand, the large  $\tilde f_d$ stable spiral region maintains stability with a big fraction of attached MPs pulling the movable MT into a very small overlap $\tilde x$, a stability with a large amount of stored free energy, i.e., low entropy. In between these two limits, for half a period MPs win the tug-of-war and in the other half PCLs win, leading to oscillation.      

\subsubsection{Experimental realizations} 
\label{sec:expt}
Experiments on wild type kinesins have provided evidence that MP- activity depends on the ambient ATP concentration.
 The intrinsic kinesin velocity, $v_0$, varies by more than one order of magnitude from $\sim 30\,$nm/s to $\sim 800\,$nm/s with change in ATP concentration from 5$\,\mu$M to 2\,mM~\cite{Schnitzer2000}.  With this change, the detachment force $f_d$ increases weakly from 1.8 to 2.4\,pN. This estimate is obtained from the measurement of run lengths of kinesins (see  Appendix-B). In terms of dimensionless parameters,  changes in   $\tilde f_d$ from 14.4 to 19.2, and $\tilde v_0$ from $\sim 1$ to 25, are accessible  varying the ATP concentration from $\mu$M to mM range. In {\em in vitro} experiments, the length of MTs may be stabilized using taxol. Fig.~\ref{fig:act_pass_pb1} suggests that carrying out such  experiments will allow one to observe all the non-equilibrium phases discussed above. Further, different types of kinesins, and kinesin mutants may allow  to probe a wider range of the  phase space in the $\tilde v_0$-$\tilde f_d$ plane.

\subsection{Both MTs movable}
\label{sec:model2}
\begin{figure}[t]
\begin{center}
\includegraphics[width=8cm]{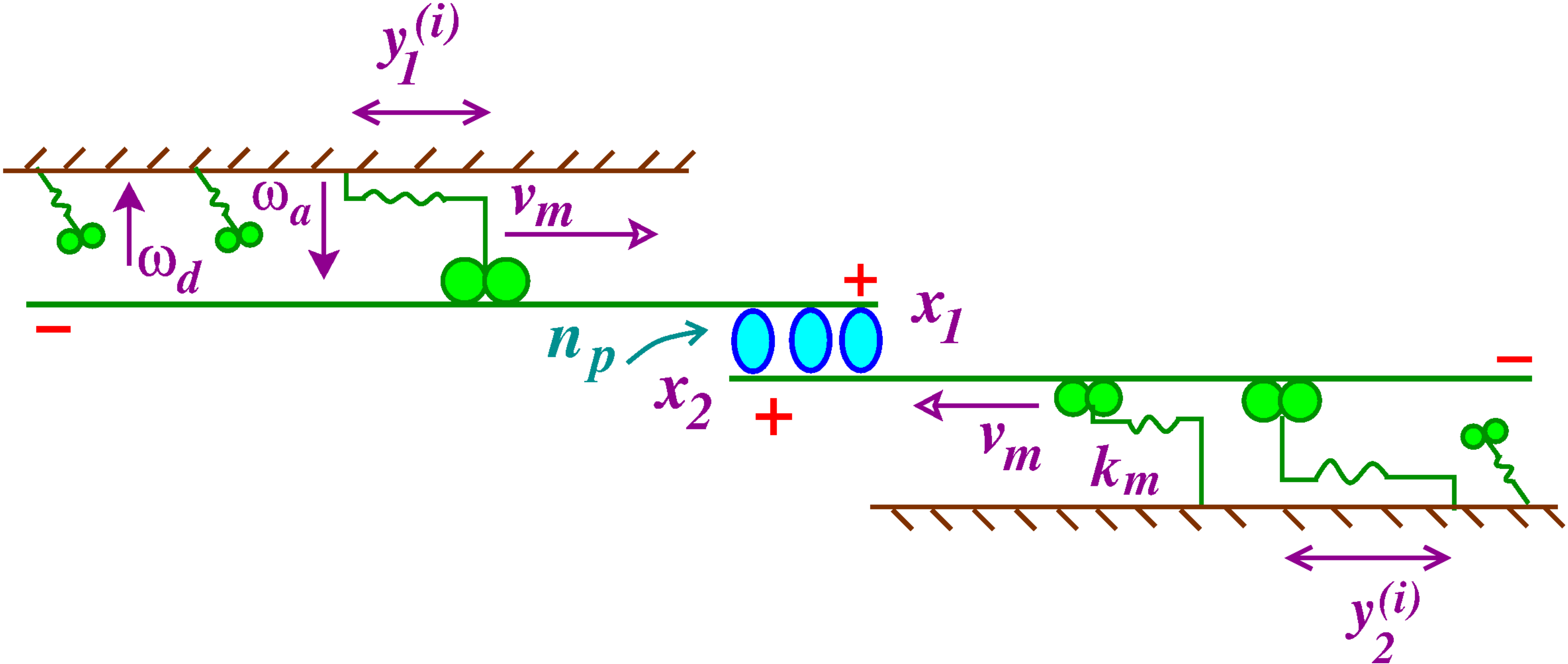}
\caption{(color online) 
Schematic diagram of the proposed {\em in vitro} experiment where two antiparallel MTs denoted by green lines and $+$, $-$ ends are driven by two sets of MPs, denoted by springs attached with double-heads.  The overlap region of the two MTs contains $n_p$ number of PCLs. The positions of the two MTs are denoted by $x_1$ and $x_2$, such that the overlap length is $x_r = x_1 - x_2$. The MPs undergo attachment detachment dynamics with rates $\w_a$ and $\w_d$ respectively. When attached, MPs move towards $+$ end of their associated MT.  MPs attached to the left (right) MT move towards right (left) with active speed $v_m$ that depends on the MP extension $y_1^{(i)}$ ($y_2^{(i)}$ ). 
As a result of active motion, MPs attached to the left (right) MT pull it towards left (right) reducing the overlap. The entropic force due to PCLs act against this pull towards increasing the overlap length $x_r$.}
\label{fig:ful_sys_mod}
\end{center}
\end{figure}

The insight gained from the above analysis allows us to extend it to a situation where both MTs are movable, as depicted in Fig.~\ref{fig:ful_sys_mod}. This geometry is closely  related to the mitotic spindle, where at the midzone antiparallel MTs are pulled apart by MPs and stabilized by PCLs. However, for simplicity and specificity, we still consider an {\em in vitro} experiment with two movable MTs on an kinesin MP assay, in the presence of Ase1 PCLs. The activity of MPs can be tuned by changing ATP concentration. 

Let $x_1$ and $x_2$ denote two tips of the antiparallel overlapping MTs. Extension of MPs associated with these two MTs are denoted by $y^{(i)}_1$ and $y^{(i)}_2$, such that the mean
extensions are $y_1 = \sum_{i=1}^{ n_{m_1} } y^{(i)}_1/n_{m_1}$ and  $y_2 = \sum_{i=1}^{n_{m_2}} y^{(i)}_2/n_{m_2}$ where $n_{m_1}$ and $n_{m_2}$ are the number of MPs attached to the left and right MTs respectively (Fig.\ref{fig:ful_sys_mod}). $N$ denotes the total number of MPs available to both the left and right MTs. The overlap length is given by $x=x_1 - x_2$. Within mean field approximation, the equations of motion in terms of dimensionless variables can be expressed as,
\begin{eqnarray}
d_\tau \tilde{x}_1 &=& \frac{1}{\exp(\zeta n_p)}\left[\frac{ n_p}{\tilde{x}_1-\tilde{x}_2} - N \tilde{n}_{m_1} \tilde{k}_m \tilde{y}_1\right], \nn\\ 
d_\tau \tilde{y}_1 &=& 
\tilde{v}_m (\tilde{k}_m \tilde{y}_1)+ d_\tau \tilde{x}_1, \nn\\ 
d_\tau \tilde{n}_{m_1} &=& (1-\tilde{n}_{m_1})\tilde{\omega} - \tilde{n}_{m_1} \exp\left[\frac{\tilde{k}_m |\tilde{y}_1|}{\tilde{f}_d}\right] ,\nn\\ 
d_\tau \tilde{x}_2 &=& \frac{1}{\exp(\zeta n_p)}\left[-\frac{ n_p}{\tilde{x}_1-\tilde{x}_2} - N \tilde{n}_{m_2} \tilde{k}_m \tilde{y}_2\right], \nn\\
d_\tau \tilde{y}_2 &=&  
-\tilde{v}_m (-\tilde{k}_m \tilde{y}_2)+ d_\tau \tilde{x}_2 ,\nn\\ 
d_\tau \tilde{n}_{m_2} &=& (1-\tilde{n}_{m_2})\tilde{\omega} - \tilde{n}_{m_2} \exp\left[\frac{\tilde{k}_m |\tilde{y}_2|}{\tilde{f}_d}\right], 
\label{eq:full_sys}
\end{eqnarray}
where, $d_\t \equiv d/d\t$, $\tilde n_{m_{1,2}} = n_{m_{1,2}} /N$, $\tilde y_{1,2} = y_{1,2}/l_0$, $\tilde x_{1,2} = x_{1,2}/l_0$, with $\{1,2\}$ denoting the two MTs and  MPs associated with them, respectively. The kinesin MPs that are attached to the right MT, move towards left, leading to a  self propelled velocity $-\tilde v_0$ and average extension $\tilde y_2 <0$. This leads to the difference between the form of 
the equations for $\tilde y_2$ with respect to that of  $\tilde y_1$. 
Note that the detachment rates enhance with the extension of MPs, irrespective of the sign of extension.

\subsubsection{Linear stability and the phase diagram}
 For a given value of parameters, the fixed point is given by $(\tilde{x}_1^0 - \tilde{x}_2^0) = {n_p}/{\tilde{n}_{m_1}^0 N \tilde{f}_s}$, $\tilde{y}_1^0 = {\tilde{f_s}} / {\tilde{k}_m}$, 
 $\tilde{y}_2^0 = -{\tilde{f_s}} / {\tilde{k}_m}$, $\tilde{n}_{m_1}^0 = \tilde{n}_{m_2}^0 = {\tilde{\omega}} / [{\tilde{\omega}+\exp({\tilde{f}_s} / {\tilde{f}_d})}]$. 
%
Linearized time evolution of small perturbations around this fixed point is given in the form of the matrix equation 
\begin{equation}
d_\tau
\begin{pmatrix}
\delta\tilde{x}_1\\
\delta\tilde{y}_1\\
\delta \tilde{n}_{m_1}\\
\delta\tilde{x}_2\\
\delta\tilde{y}_2\\
\delta \tilde{n}_{m_2}
\end{pmatrix}=
\begin{pmatrix}
b_{11} & b_{12} & b_{13} & -b_{11} & 0 & 0\\
b_{11} & b_{22} & b_{13} & -b_{11} & 0 & 0 \\
0 & b_{32} & b_{33} & 0 & 0 & 0\\
-b_{11} & 0 & 0 & b_{11} & b_{12} & -b_{13}\\
-b_{11} & 0 & 0 & b_{11} & b_{22} & -b_{13}\\
0 & 0 & 0 & 0 & -b_{32} & b_{33}
\end{pmatrix}
\begin{pmatrix}
\delta\tilde{x}_1 \\
\delta\tilde{y}_1\\
\delta \tilde{n}_{m_1}\\
\delta\tilde{x}_2\\
\delta\tilde{y}_2\\
\delta \tilde{n}_{m_2}\\
\end{pmatrix}
\label{eq:six_by_six}
\end{equation}
where,
\begin{align}
& b_{11} = -{(N \tilde{n}_{m_1}^0 \tilde{f}_s)^2} / { n_p \exp(\zeta n_p)},
 b_{12} =  -{N \tilde{n}_{m_1}^0 \tilde{k}_m} / {\exp(\zeta n_p)}, \nn\\
& b_{13} =  -\frac{N \tilde{f}_s}{\exp(\zeta n_p)}, a_{22} =  -\left[\frac{\tilde{v}_o \tilde{k}_m}{\tilde{f}_s} + \frac{N \tilde{n}_{m_1}^0 \tilde{k}_m}{\exp(\zeta n_p)}\right],\nn\\ 
& b_{32} =  -{\tilde{n}_{m_1}^0 \tilde{k}_m} / {\tilde{f}_d} \exp({\tilde{f}_s} / {\tilde{f}_d}), 
b_{33} =  -\tilde{\omega}-\exp({\tilde{f}_s} / {\tilde{f}_d}).\nn
\end{align}

Let us define a set of new variables using the following linear transformations $\bar x = (\tilde x_1 + \tilde x_2), x_r = (\tilde x_1 - \tilde x_2), \bar y = (\tilde y_1 - \tilde y_2), y_{r} = (\tilde y_1 + \tilde y_2), \bar n_m = (\tilde n_{m_1} + \tilde n_{m_2}), n_m^{r} = (\tilde n_{m_1} - \tilde n_{m_2})$.  In these definitions, some of the variables has clear physical meaning and interest, e.g.,  $x_r$ is the overlap length of the MTs, $\bar y$ denotes the total mean extension of the MPs generating pulling force on the overlap, and $\bar n_m$ denotes the total number of motors generating the pull. The center of mass of the MT pair is denoted by $\bar x/2$. In terms of these new variables,  Eq.(\ref{eq:six_by_six}) breaks down into a block diagonal form with each block being independent of the other.

The dynamics of the overlap length $x_r$ depends on the force due to MPs via the mean MP 
extension $\bar y$ and the total number of MPs in the attached state $\bar n_m$, and is governed by,
\begin{equation}
d_\tau
\begin{pmatrix}
\delta {x}_r\\
\delta \bar {y}\\
\delta \bar {n}_m\\
\end{pmatrix}=
\begin{pmatrix}
2b_{11} & b_{12} & b_{13}\\
2b_{11} & b_{22} & b_{13}\\
0 & b_{32} & b_{33}\\
\end{pmatrix}
\begin{pmatrix}
\delta {x}_r\\
\delta \bar {y}\\
\delta \bar {n}_m\\
\end{pmatrix}.
\label{eq:three_by_three}
\end{equation}
The stability matrix in the above equation,  $\utilde b$ , has exactly the 
same structure as {$\utilde a$} in the previous section, only difference being that in  $\utilde b$  the element $2 b_{11}$ has
replaced the element $a_{11}$ of the matrix {$\utilde a$}. 

The relative variables $y_r$ and $n_m^r$, gets decoupled and can be expressed in terms of the stability matrix $\utilde b'$ such that,
\begin{equation}
d_\tau
\begin{pmatrix}
\delta{y}_r\\
\delta{n}^{r}_m\\
\end{pmatrix}=
\begin{pmatrix}
b_{22} & b_{13}\\
b_{32} & b_{33}\\
\end{pmatrix}
\begin{pmatrix}
\delta{y}_r \\
\delta{n}^{r}_m\\
\end{pmatrix}.
\label{eq:two_by_two}
\end{equation}

Finally, while the center of mass of the overlapping MT pair does not impact any other variable, it itself evolves as
\begin{equation}
d_{\tau}\bar x = b_{12}y_r + b_{13} n_m^r\label{eq:linear}.
\end{equation}
%

Eq.(\ref{eq:three_by_three}) is completely independent of the dynamics of $\bar x$, $y_{r}$ and $n_m^r$. Thus stability of Eq.(\ref{eq:two_by_two}) and Eq.(\ref{eq:linear}) does not affect the stability of Eq.(\ref{eq:three_by_three}) involving the overlap $x_r$. Consequently, diagonalizing the $3\times 3$ stability matrix  $\utilde b$  we get the characteristic equation
\begin{eqnarray}
\lambda^3 + A' \lambda^2 + B' \lambda + C' &=& 0, \nn
\end{eqnarray}
%
where, $A' = - {\rm Tr}$( $\utilde b$ ), $B' = \hf (b_{ii} b_{jj} - b_{ij} b_{ji})$ and $C' = -$det( $\utilde b$ ) are equivalent to the coefficients $A,\, B,\, C$ in the previous section~\footnote{
The full expressions of the coefficients in terms of the elements of the stability matrix   $\utilde b$   read
\unexpanded{\begin{eqnarray}
A' &=& \frac{2(N \tilde{n}_{m_1}^0 \tilde{f}_s)^2}{ n_p \exp(\zeta n_p)} + \left[\frac{\tilde{v}_o \tilde{k}_m}{\tilde{f}_s} + \frac{N \tilde{n}_{m_1}^0 \tilde{k}_m}{\exp(\zeta n_p)}\right] + \tilde{\omega} + \exp\left[\frac{\tilde{f}_s}{\tilde{f}_d}\right], 
\label{eq:six_six_co_p}\nn\\
B' &=& \frac{2(N \tilde{n}_{m_1}^0 \tilde{f}_s)^2}{ n_p \exp(\zeta n_p)} \left[\frac{\tilde{v}_o \tilde{k}_m}{\tilde{f}_s} + \tilde{\omega} + \exp\left[\frac{\tilde{f}_s}{\tilde{f}_d}\right]\right] \nonumber\\&& + \frac{\tilde{v}_o \tilde{k}_m}{\tilde{f}_s} \left[\tilde{\omega} + \exp\left[\frac{\tilde{f}_s}{\tilde{f}_d}\right]\right] \nonumber\\&& + \frac{N \tilde{n}_{m_1}^0 \tilde{k}_m}{\exp(\zeta n_p)}\left[\tilde{\omega} + \left(1-\frac{\tilde{f}_s}{\tilde{f}_d}\right)\exp\left[\frac{\tilde{f}_s}{\tilde{f}_d}\right]\right], 
\label{eq:six_six_co_q} \nn\\
C' &=& \left(\frac{2 (N \tilde{n}_{m_1}^0 \tilde{f}_s)^2}{ n_p \exp(\zeta n_p)}\right)\left(\frac{\tilde{v}_o \tilde{k}_m}{\tilde{f}_s}\right) \left(\tilde{\omega}+\exp\left[\frac{\tilde{f}_s}{\tilde{f}_d}\right]\right).\nn 
\label{eq:six_six_co_r}
\end{eqnarray}
}}.
%
\begin{figure}[t]
\begin{center}
\includegraphics[width=8cm]{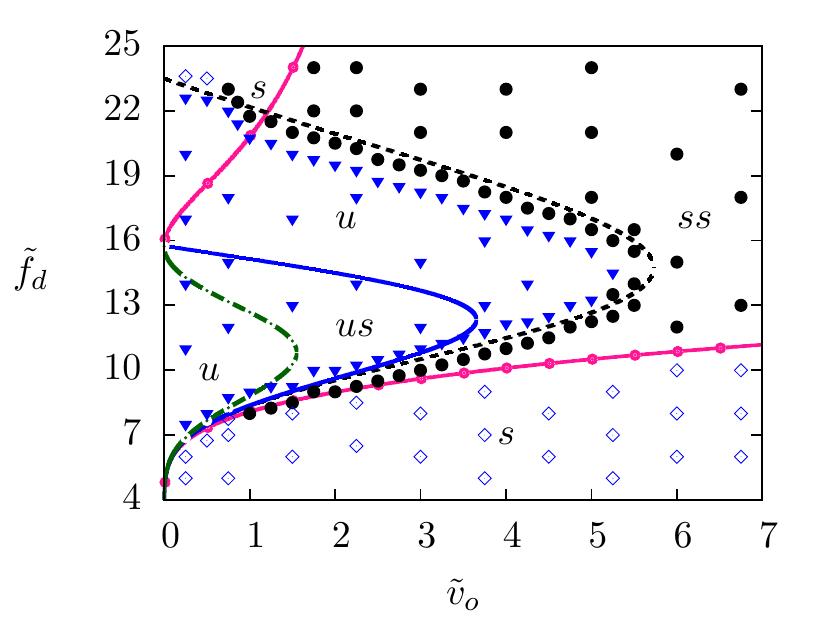} 
\caption{ (color online) 
Phase diagram when both the MTs of overlapping pair are considered movable, in the $\tilde v_0-\tilde{f}_d$ plane. The parameter values used $\tilde k_m = 450$, $\tilde\omega = 20$, $\tilde{f}_s = 60$, $N = 5$ and $n_p = 15$. Linear stability estimate of different phases are denoted by $s$: { linearly stable} phase, $u$: { linearly unstable} phase, $us$: { unstable spiral } phase and $ss$: { stable spiral } phase, while the lines show the corresponding phase boundaries.  
The points denote non-linear dynamics estimate of the phases obtained from numerical integration of Eq.(\ref{eq:full_sys}):  stable {\color{blue} $\diamond$}, unstable {\color{blue} $\blacktriangledown$},  
stable spiral \tikzcircle[black, fill=black]{2pt}. Note that the linear instability due to Eq.(\ref{eq:two_by_two}) destabilizes all other possible phase behavior within the dashed black line. 
}
\label{fig:act_pass_pb2}
\end{center}
\end{figure}
%
{Analyzing properties of $\l$ in the same manner as in Sec.~\ref{sec:model1}, we obtain the phase boundaries presented by lines in Fig.(\ref{fig:act_pass_pb2}). As expected, this leads to a prediction of the same qualitative features of overlap as in Fig.(\ref{fig:act_pass_pb1}), showing stable node, stable spiral, unstable node and unstable spiral phases.

The relative quantities $y_r,\, {n}^{r}_m$ may be excited using an asymmetric perturbation in $y_{1,2}$ and $n_{m_{1,2}}$. Their decoupled dynamics given in Eq.~(\ref{eq:two_by_two}) can be analyzed by diagonalizing the stability matrix  $\utilde b'$   to obtain the characteristic equation $\lambda^2 - p \lambda + q = 0,$ where, $p=$Tr.(\,$\utilde b'$ ) $= b_{22} + b_{33}$, and $q =$ det( $\utilde b'$ ) $= b_{22} b_{33} - b_{13} b_{32}$.  The eigenvalues $\lambda_{1,2} =(p \pm \sqrt{p^2-4q})/2$, with $p<0$ while $q$ may change sign. With change in parameters, if $q$ becomes negative, $y_r$ and $n_m^r$ gets unstable. Thus $q=0$ denotes the phase boundary of instability, which is shown in Fig.(\ref{fig:act_pass_pb2}) by the dashed black line.
}
One may find oscillations in this asymmetric perturbation if the discriminant $\D = p^2-4q$ is negative, however, $\D = (b_{22} - b_{33})^2 + 4b_{13}b_{32} > 0$, discarding any possibility of oscillation in this  mode.

A general perturbation contains all possible components, symmetric and asymmetric. Although these components are linearly decoupled, as shown above, instability in one can destabilize all other quantities. Numerical integration of Eq.(\ref{eq:full_sys}) shows non-equilibrium phases characterized by stability, stable spiral behavior, and instability in the MT overlap. 
The linear instability in asymmetric fluctuations characterized by $y_r$, $n_m^r$ destabilzes any possibility of limit cycle oscillations in overlap $x_r$ within the region enclosed by the dashed black line in Fig.~\ref{fig:act_pass_pb2}. This behavior is unlike that in Fig.~\ref{fig:act_pass_pb1}, overriding the similarities of ($x_r$, $\bar y$, $\bar n_m$) mode with one movable MT  case. The phase diagram, Fig.~\ref{fig:act_pass_pb2}, shows that the overlap between MTs has a stable region, either linear or oscillatory, denoted by a stable node or spiral respectively. The MTs may also show linear instability towards complete overlap. In stark contrast to Fig.~\ref{fig:act_pass_pb1}, the stable limit cycle oscillation phase is completely ruled out. The fact that when both the MTs in an overlapping pair are movable, the competition between MPs and PCLs lead to stability of partial overlap or instability towards complete overlap, is consistent with the predictions of Ref.~\cite{Johann2015}. In Fig.~\ref{fig:phdia2}($b$) of  Appendix-D, we show a phase diagram for another set of values of $n_p$ and $N$, adding to the understanding of overall structure of the phase boundaries.

\section{Discussion}
We considered two related systems, analyzing competition between active and passive forces on the overlap of filaments cross-linked by PCLs and driven by MPs.  PCLs impact this dynamics via entropic force generation and enhancement of viscous friction between the overlapping pair. Within a mean field approximation, the motion of MTs and associated MPs in the presence of PCLs is described by a set of coupled non-linear differential equations. The resultant dynamics was analyzed within linear stability and using numerical integration. This showed non-equilibrium phases characterized by stability, via either stable node or stable spiral, instability towards complete overlap, and stable limit cycle oscillations. For a given number of PCLs, the transitions between these non-equilibrium phases can be controlled by MP activity that depends on ambient ATP concentration. A dramatic change in the dynamics occurs depending on whether one of the partner MT of an overlapping pair is held fixed, or both the MTs are movable.  

If one MT is immobilized, the overlap dynamics can give rise to a stable limit cycle oscillatory phase that occurs via a supercritical Hopf bifurcation, where the oscillations are maintained by the inherent non-linearity, cooperativity through load sharing, and a phase lag between the mean MP extension and the evolution of overlap.  Earlier studies mainly focussed  on the impact of active mechanism due to MPs on mitotic spindle. In contrast,  we have shown that the PCLs present at the overlap play a crucial role in controlling the non-equilibrium phases. In particular, PCLs control phase boundaries and the oscillation frequency at the onset of limit cycle. Within the stable limit cycle phase, the oscillation amplitude of overlap stays within tens of nm and the frequency varies by one order of magnitude between 0.1 to 1 Hz with the change in activity parameters.  This frequency range is within that of various other cellular processes, and may potentially interfere with them. 

The possibility of having a stable limit cycle oscillation phase disappears as the fixed partner of an anti-parallel MT pair is allowed to move. Any asymmetric perturbation between the two MTs grows to destabilize the possibility of stable oscillations. The cell may control the switching between a stable limit cycle oscillatory phase and non-oscillatory phase, by fixing or releasing one of the partner MTs of the overlapping pair to or from the cytoskeletal matrix using molecular bonds. The non-equilibrium phases might be tuned by the cell  also by controlling the local ATP concentration,  and changing the number of PCLs present in the overlap. Our work is thus directly relevant for understanding the impact of PCLs in the stability of mitotic spindle structure. We used parameterization considering MTs under the influence of kinesin MPs and Ase1 PCLs. This makes our predictions amenable to quantitative comparison with  {\em in vitro} experiments. 

We performed all the analysis within a mean field approximation ignoring stochasticity. It remains to be seen how consideration of passive and active noise impacts the mean field phase diagram. It will be interesting to study the influence of PCL turnover on the overlap dynamics.

\acknowledgements

DC thanks  Amitabha Nandi  and Chaitra Hegde for  discussions, and  in particular Pieter Rein ten Wolde for his useful comments and suggestions.
D.C. acknowledges SERB, India for financial support through grant numbers EMR/2016/0001454, and EMR/2014/000791.  I.P. acknowledges MINECO and DURSI for financial support
under projects FIS2015-67837- P and 2014SGR-922, respectively. 

\appendix

\section{Dependence of viscous friction on PCLs}
\label{ap_friction}
\begin{figure}[t]
\begin{center}
\includegraphics[width=8cm]{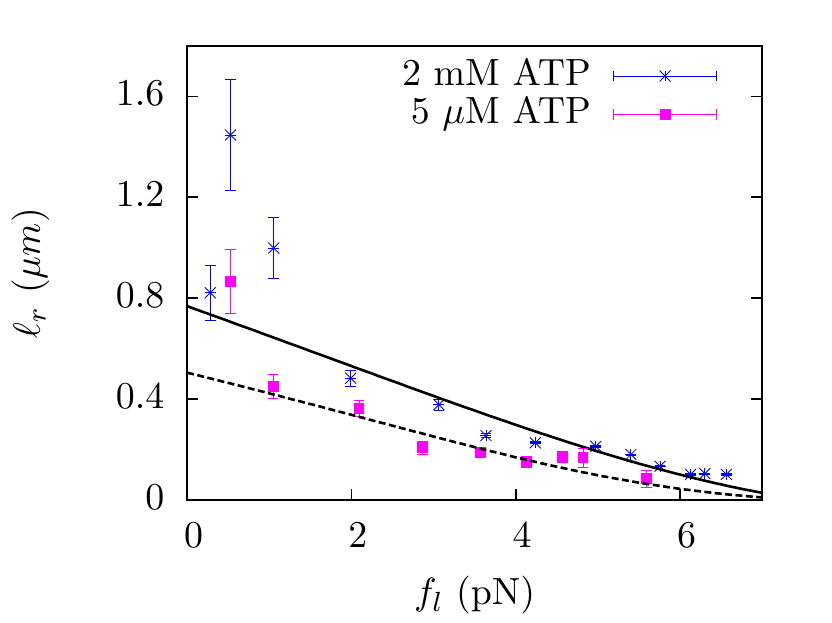}
\caption{ (color online) Experimental data for kinesin extracted from  Fig.(3b)  of Ref.~\cite{Schnitzer2000} at two different ATP concentrations. The lines show fit to the functional form 
$v_0(1-{f_l}/{f_s})\, \bar\k \w_a\, [\w_a + \w_d \exp( {f_l} / {f_d}) ]^{-1}$. Here, independent estimates of active velocities $v_0 = 0.8\,\mu$m/s and $0.05\,\mu$m/s at respective ATP concentrations of 2mM and 5$\,\mu$M are kept fixed~\cite{Chaudhuri2016,Schnitzer2000}. The fitting procedure gives $f_d = 2.4\,$pN, and $1.8\,$pN along with $\bar\k=1,\,10\,$s for the two ATP concentrations, respectively.  Here $f_l$ denotes the load force, $f_s = 7.5$ pN, $\omega_a = 20/s$, $\omega_d = 1/s$. }
\label{findfd}
\end{center}
\end{figure}
The mechanism of generation of viscous friction due to the PCLs may be described as follows. 
In presence of PCLs, a MT must overcome an energy barrier in order to move with respect to the other MT. Assume that the PCLs are extensible springs with spring constant $k_p$ with the two ends of a PCL attached to two MTs. Let the bottom end of one PCL move by a distance $\d$ to the right, keeping the other end fixed. This will pull the top MT to the right, while rest of the $n_p-1$ PCLs will pull it to the left bringing the system to a new force balance, with increased potential energy. The relative motion of MTs is mediated by hopping of PCLs. The energy reaches its maximum when equal number of PCLs, $n_p/2$, pull to the right and left. This  transition state energy is $ 2\times\, \hf k_p \d^2 \times (n_p/2)$. The top MT will slide to right with respect to the bottom MT, once this energy barrier is crossed~\cite{Lansky2015}. The rate of this barrier crossing $\sim \exp(- k_p \d^2 n_p/2 \kb T)$ gives the MT diffusivity $D_{MT} = \kb T/\g_f$, leading to an exponential increase of viscous friction with the number of PCLs $\g_f = \g \exp(\zeta n_p)$ where $\zeta = k_p \d^2/2 \kb T$. 
A comparison with experiments gives $\g = 8.93 \times 10^2\, \kb$T-s/$\mu$m$^2$ and $\zeta = 0.22$~\cite{Lansky2015}, assuming fluorescent intensity from Ase1 as equivalent to $n_p$.  
Within this picture, we ignore the viscous drag due to the ambient fluid as it turns out to be orders of magnitude smaller than that  typical magnitudes of the rest of the forces involved. For example, the viscous friction felt by a MT of $10\,\mu$m length in water (of viscosity 0.001 Pa-s) is $\sim 10^{-3}\, \kb$T-s/$\mu {\rm m}^2$, five orders of magnitude smaller than $\g_f$. 

\begin{figure*}[htb]
\begin{center}
\includegraphics[width=8cm]{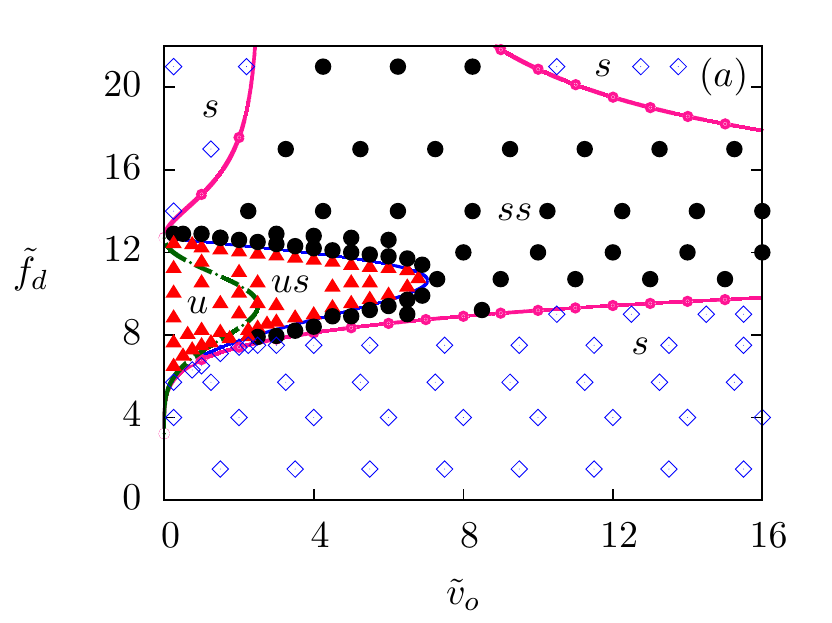} 
\includegraphics[width=8cm]{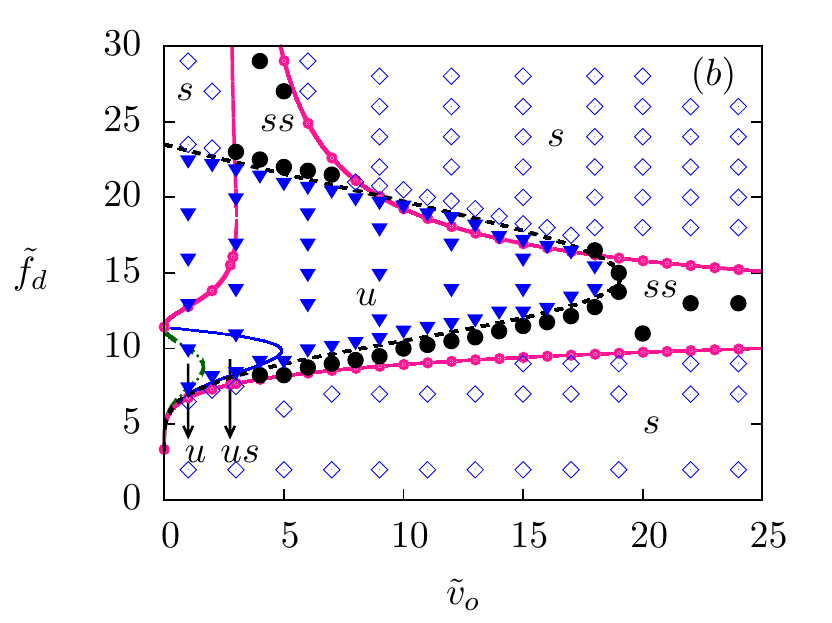} 
\caption{ (color online) 
Phase diagrams in $\tilde v_0-\tilde{f}_d$ plane with one movable MT  ($a$), and both MTs movable  ($b$). The parameter values used $\tilde k_m = 450$, $\tilde\omega = 20$, $\tilde{f}_s = 60$, 
$N = 50$ and $n_p = 20$. The region $s$ indicates { linearly stable} phase, region $u$ indicates { linearly unstable} phase, region $us$ indicates { unstable spiral } phase and region $ss$ indicates { stable spiral } phase corresponding to linear stability analysis. The points denote corresponding non-linear dynamics estimate obtained from numerical integration:  stable {\color{blue} $\diamond$}, unstable {\color{blue} $\blacktriangledown$}, stable spiral \tikzcircle[black, fill=black]{2pt} and stable limit cycle {\color{red} $\blacktriangle$}. Note the stark contrast between the two phase diagrams, while diagram ($a$) does not have any  unstable {\color{blue} $\blacktriangledown$} region, diagram ($b$) is devoid of stable limit cycle {\color{red} $\blacktriangle$}. 
}
\label{fig:phdia2}
\end{center}
\end{figure*}

\section{Dependence of $\tilde f_d$ on ATP concentration}
\label{app_fd_theory}

A free MP, when attached  to a conjugate filament, moves with velocity $v_0$ dependent on ATP concentration. 
The time-scale over which a MP remains associated to a MT and moves actively is proportional to the duty ratio 
$\o = \w_a/[\w_a + \w_d \exp(f_l/f_d)]$. 
In the presence of resistive load $f_l$, thus, the run length of a kinesin molecule over MT is given by $\ell_r = \bar\k v_m(f_l) \o$, 
where, $\bar\k$ is a proportionality constant having the dimension of time. Measurements of such run-lengths for kinesins at two different ATP concentrations were presented in Ref.~\cite{Schnitzer2000}. We use this data to extract the detachment force $f_d$ (see Fig.~\ref{findfd}). Fitting the data to $\ell_r = \bar\k v_m(f_l) \o$ within the linear force-velocity regime,  shows that $f_d$ does not vary much, although values of $v_0$ strongly depend on the ATP concentration. Fixing values of $v_0$ from independent measurements of Ref.~\cite{Schnitzer2000},
we find that  $f_d = 2.4\,$pN for 2mM ATP concentration, and $f_d = 1.8\,$pN for 5$\mu$M ATP concentration~(Fig~\ref{findfd}). 

\section{Calculating phase boundary}
\label{sec:ph_bc}
At the phase boundary, with two degenerate roots at $\l_m =  -\frac{A}{3} + \frac{1}{3}\sqrt{A^2 - 3B}$, the characteristic equation is
$(\lambda + \lambda_1)(\lambda - \lambda_m)^2 = 0$.
Comparing it with Eq.(\ref{Eq:cubic_act_pass01}) one gets
\begin{eqnarray}
A &=& -(\lambda_1 + 2\lambda_m) \label{eq:a} \\
B &=& \lambda_m^2 + 2\lambda_1\lambda_m \label{eq:b} \\
C &=& \lambda_1\lambda_m^2. \label{eq:c} 
\end{eqnarray}
Eq.s (\ref{eq:a}) and (\ref{eq:b}) are satisfied by $\lambda_m$ and
\begin{eqnarray}
\lambda_1 &=& \frac{1}{3}A + \frac{2}{3}\sqrt{A^2 -3B}. \nn
\end{eqnarray}
As a result Eq.(\ref{eq:c}) gives the  equation of phase boundaries 
\begin{equation}
C = \bigg[\frac{1}{3}A + \frac{2}{3}\sqrt{A^2 -3B}\bigg]\bigg[-\frac{1}{3}A + \frac{1}{3}\sqrt{A^2 -3B}\bigg]^2,
\label{eq:pb_ReIm}
\end{equation}
with $B\textgreater0$ denoting the boundary between $ss$ and $s$ phases while $B\textless0$ gives the boundary between $us$ and $u$ phase. 

\section{Phase diagrams}
\label{app_phdia}
In Fig.~\ref{fig:phdia2}($a$) and ($b$), we show the phase diagrams corresponding to the two models discussed in Sec.~\ref{sec:model1} and Sec.~\ref{sec:model2} for different number of PCLs $n_p=20$ and total motor proteins $N=50$. The overall structure of the phase diagrams remain unaltered, however, the appearance of particular phases does depend on the number of PCLs, $n_p$. The upper branch of the linear stability line between the stable ($s$) and the stable spiral ($ss$) phases are clearly visible in these diagrams.

\section{Time scales near phase boundary}
\label{ap_divergence}
\begin{figure}[h]
\begin{center}
\includegraphics[width=8cm]{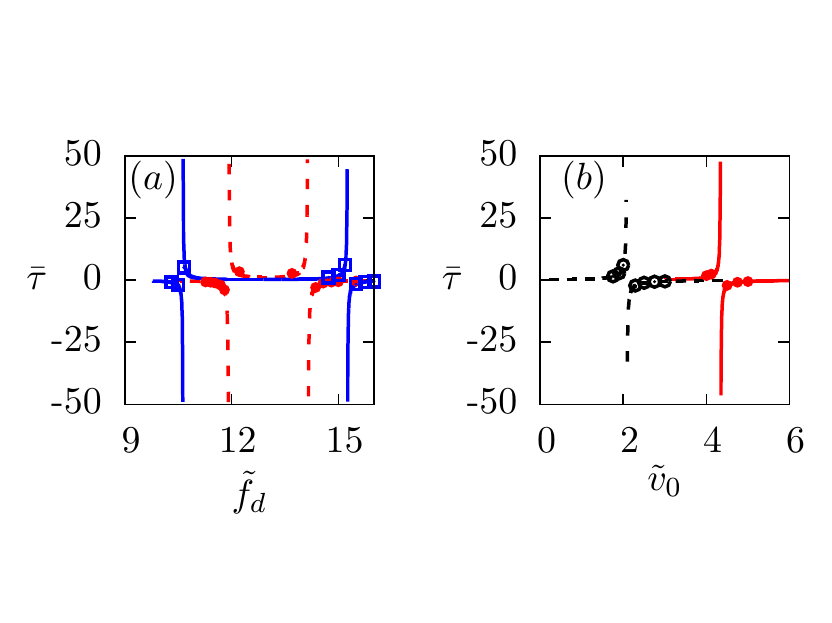}
\caption{ (color online) 
Time scales obtained from mean field analysis of stability and instability with ($a$) change in $\tilde f_d$ at $\tilde v_0=3$ (blue solid line), and $\tilde v_0=4$ (red dashed line),
($b$) change in $\tilde v_0$ at $\tilde f_d=13$  (red solid) and $\tilde f_d=16$ (black dashed). The points denote the same obtained from the numerical solution of the non-linear dynamical equations, Eq.(\ref{eq:dyn1}).
}
\label{fig:tscale}
\end{center}
\end{figure}

The linear stability analysis gives an estimate of the time-scale $\bar \t$ of growth or decay of the perturbation as $\exp(t/\bar \t)$ near the onset of instability, be it linear or oscillatory.   
Here $\bar \t >0$ denotes time required for instability to grow, and $\bar \t <0$ denotes time required for a perturbation to die out within a stable phase.   
Fig.\ref{fig:tscale} shows these estimates around phase transitions in Fig.(\ref{fig:act_pass_pb1}). Magnitude of both the time scales are small deep inside a given phase, and shows signature of divergence at the phase boundaries. 

\end{document}